\documentstyle[12pt,amstex,amssymb]{article}

\def \ub {\underline}
\def \id {1\!\!{\rm I}}

\def \ca {({\cal A})}
\def \a {{\cal A}}
\def \ce {({\cal E})}
\def \e {{\cal E}}
\def \oc {\Omega^{\bf \centerdot}}
\def \toc {\tilde\Omega^{\,^{\bf{\cdot}}}}
\def\cint{\ifinner{\int\hspace*{-0.35cm}-}\else{\int\hspace*{-0.4cm}-}\fi}

\numberwithin{equation}{section}

\begin{document}


\rightline{\bf ETH-TH/95-3}

\vspace{2cm}

\begin{center}
\section*{The Gravitational Sector in the Connes-Lott\\
Formulation of the Standard Model}
\end{center}

\vspace{2truecm}

\begin{center}{\bf A.H. Chamseddine, \ \ \ J. Fr\"ohlich, \ \ \
  O. Grandjean}
\end{center}

\vspace{1truecm}

\begin{center}
Theoretical Physics, ETH-H\"onggerberg\\
CH--8093 Z\"urich, Switzerland
\end{center}

\vspace{4truecm}




\noindent {\bf Abstract }\ \ We study the Riemannian aspect and the
Hilbert-Einstein gravitational
action of the non-commutative geometry underlying the Connes-Lott
construction of the action functional of the standard model. This
geometry involves a two-sheeted, Euclidian space-time. We show that if
we require the space of forms to be locally isotropic and the Higgs
scalar to be dynamical, then the Riemannian metrics on the two sheets
of Euclidian space-time must be identical. We also show that the
distance function between the two sheets is determined by a single,
real scalar field whose VEV sets the weak scale.


\baselineskip=.7truecm

\vfill\eject

\section{Introduction}

Among recent ideas on the structure of space-time at short
distance scales the proposal that space-time is an aspect of a
non-commutative (metric) space has the appealing feature that it
allows one to develop a natural geometric setting for the standard
model [1,2]. In particular, the Higgs sector finds a natural
geometrical interpretation.

The non-commutative space underlying the standard model is built over
a Euclidian space-time consisting of two copies of Euclidian space
${\Bbb E}^4$. The weak scale turns out to be given by the inverse of
the distance between the two copies. If one wants to incorporate
gravitational interactions (at least at the classical level) into this
formulation of the standard model -- as one should do, in principle --
the two copies of ${\Bbb E}^4$ must be replaced by a pair of two
diffeomorphic Riemannian manifolds which, in general, will have
non-vanishing curvatures, and the distance between these two manifolds
will be described by a position-dependent, real (more precisely,
positive) field, instead of a constant. In ref.~[3], we have developed
a notion of non-commutative {\it Riemannian} geometry, following the
ideas in [2], and we have studied a simple example built upon a
two-sheeted Riemannian space $M_4 \times {\Bbb Z}_2$. The metric and
the Levi-Civita connection on the analogue of the cotangent bundle
then depend on the choice of a Riemannian metric on $M_4$ and of a
real scalar field determining the distance between the two sheets. One
can then consider the Hilbert-Einstein action in this example and
finds that, besides the usual term proportional to the integral of the
scalar curvature over $M_4$, it contains a kinetic term for the scalar
field minimally coupled to the metric on $M_4$.

An alternative approach to studying the gravitational sector of the
standard model has been proposed [4-7] in which the analogue of the
Hilbert-Einstein action is defined as the ``Wodzicki residue'' of the
inverse of the square of a covariant Dirac operator on $M_4\times{\Bbb
  Z}_2$, or, equivalently, as the $a_2$-coefficient in the expansion
of the trace of the heat kernel associated with the square of the
Dirac operator. This definition yields a Hilbert-Einstein action which
is the sum of the usual term proportional to the scalar curvature and
a term proportional to the square of the scalar field. But there is no
kinetic energy term for the scalar field, and thus this field does not
propagate. The problem with this approach is that when one couples
matter to gravity [8] and then eliminates the non-dynamical scalar
field, using its equation of motion, one ends up with a complicated
non-linear sigma model involving the matter fields that looks rather
meaningless. This problem can be avoided by defining the gravitational
action in terms of the $a_4$-coefficient (instead of the
$a_2$-coefficient) in the expansion of the trace of the heat kernel of
the square of the Dirac operator. One then arrives at a Weyl-invariant
action functional for gravity, and the scalar field has a kinetic term
that enables it to propagate [7]. But now the problem arises to see
whether there is a mechanism to avoid the spin-2 ghost mode in the
metric; a problem that is unsolved.

All these difficulties compell us to return to the strategy proposed
in [3] and construct the gravitational action for the standard model
in a more conceptual way, using the tools of non-commutative
Riemannian geometry. In Sect.~2, we review very briefly Connes'
concept of non-commutative geometry and some basic tools of
non-commutative Riemannian geometry. Readers who are less
mathematically inclined can proceed directly to Section~3. In Sect.~3,
we consider the Connes-Lott construction of the action functional of
the standard model. We prove that the requirements that there be a
non-trivial Higgs field and that the space of differential forms be
locally isotropic imply that the Riemannian metrics on the two sheets
of Euclidian space-time must be {\it identical}. We also show that a
{\it single}, real scalar field, setting the weak scale, appears as
additional gravitational degrees of freedom, besides those described
by the standard space-time metric. We determine the Hilbert-Einstein
action and eliminate auxiliary fields by using their equations of
motion.

\vspace{1cm}

\section{Some notions and tools of non-commutative\hfill\break
Riemannian geometry}

The structure of a topological manifold, $M$, is coded into the
structure of the abelian algebra, ${\cal A}_M$, of complex-valued,
continuous functions on $M$. The algebra ${\cal A}_M$ is a \break
$^*$-algebra,
the $^*$ operation being given by complex conjugation of functions. The
manifold $M$ can be viewed as the space of characters of ${\cal
  A}_M$. If $M$ is compact ${\cal A}_M$ is unital, i.e., it contains
an identity 1, the constant function equal to 1 on $M$. Connes'
proposal is to define a compact, non-commutative space in terms of a
unital, non-abelian $^*$-algebra ${\cal A}$, the ``algebra of functions
on the non-commutative space''; [2].
A non-commutative space in this sense represents relatively little
mathematical structure. In order to develop a differential geometry of
non-commutative spaces, one must add more structure; (see [1,2] and
[3,9]). Before we describe what structure to add, we briefly review
what can be developed from the data introduced so far: a unital
$^*$-algebra ${\cal A}$.

Connes [2] defines a ${\Bbb Z}$-graded differential unital algebra of
universal forms, $\oc ({\cal A})$, over ${\cal A}$. The
algebra $\oc ({\cal A})$ is generated by elements $a\in{\cal
  A}$ of degree 0 and elements $da$, $a\in{\cal A}$, of degree 1, with
relations $d(a+b)=da+db$, $d(ab)=da\,b+a\,db$ (Leibniz rule), for
$a,b$ in ${\cal A}$, and $d1=0$. An element $\alpha\in\oc
({\cal A})$ is said to have degree $n$ if it has the form
\begin{equation}
\alpha\;=\;\sum_j a_j^0\;da_j^1 \dots da_j^n\quad , \; a_j^i \in {\cal
  A}.
\end{equation}
Let $\Omega^n ({\cal A})$ be the vector space of elements of degree
$n$. Using the Leibniz rule one verifies that $\oc ({\cal A})
= \displaystyle\mathop{\oplus}_{n=0}^\infty\;\Omega^n ({\cal A})$,
with $\Omega^0 ({\cal A}) = {\cal A}$, and $\Omega^i ({\cal A})
\Omega^j ({\cal A}) \subset \Omega^{i+j} ({\cal A})$. The differential
$d$ on $\oc ({\cal A})$ is a linear map of degree 1 defined
by
\[
d (a_0 da_1 \cdots da_n)\;=\;da_0 \,da_1 \cdots da_n\;,\quad a_i \in
{\cal A}.
\]
The identity of $\oc ({\cal A})$ is given by $1\in {\cal A} =
\Omega^0 ({\cal A})$. In fact, $\oc ({\cal A})$ becomes a
$^*$-algebra by defining
\begin{equation}
(da)^*\;=\;-\,da^*\;,\quad
(\alpha\beta)^*\;=\;\beta^*\,\alpha^*\;,\quad
a\,\in\,{\cal A},\;,\quad
\alpha,\beta \,\in\,\oc \ca .
\end{equation}
The cohomology of $\oc\ca $ is trivial.

The $K$-theory of the algebra ${\cal A}$ is the
study of ``vector bundles over the non-commut\-at\-ive space described by
${\cal A}$''. Inspired by Swan's theorem for vector bundles over
compact manifolds, [10], one defines (the space of sections of) a
vector bundle, ${\cal E}$, over the non-commutative space described by
${\cal A}$ as a {\it finitely generated, projective left ${\cal
    A}$-module}, [2]. A {\it connection} $\nabla$ on ${\cal E}$ is
defined to be a linear map
\begin{equation}
\nabla\;:\;{\cal E}\;\longrightarrow\;\Omega^1 \ca
\displaystyle\mathop{\otimes}_{{\cal A}}\;{\cal E}
\end{equation}
such that, for any $a\in{\cal A}$ and $s\in{\cal E}$,
\begin{equation}
\nabla (as)\;=\;da\,\otimes\,s\,+\,a\,\nabla s .
\end{equation}
Given ${\cal E}$, we define
$\oc \ce$ to be the graded left $\oc \ca$-module
given by
\begin{equation}
\oc \ce\;=\;\oc \ca \,\displaystyle\mathop{\otimes}_{\a}\,\e,
\end{equation}
where we are using that $\Omega^\centerdot \ca$ is a left and right
$\a$-module.
One calls $\oc \ce$ the space of $\e$-valued, universal
forms. A connection $\nabla$ on $\e$ extends uniquely to a linear map
of degree one
\[
\nabla\;:\;\oc \ce\;\longrightarrow\;\oc \ce
\]
with
\begin{equation}
\nabla
(\alpha\sigma)\;=\;d\alpha\sigma\,+\,(-1)^{deg\;\alpha}\;\alpha\,\nabla\,\sigma
\end{equation}
for any homogeneous $\alpha\in\oc\ca$ and any
$\sigma\in\oc \ce$. This observation enables one to define
the {\it curvature} of a connection $\nabla$ by setting
\begin{equation}
R(\nabla)\;:=\;-\,\nabla^2\;:\;\e\,\longrightarrow\,\Omega^2 \ca
\,\displaystyle\mathop{\otimes}_{\a}\,\e .
\end{equation}
One easily checks (using that $d^2=0)$ that
\begin{equation}
R (\nabla)\,(as)\;=\;a\;R (\nabla) s
\end{equation}
for any $a\in \a$ and any $s\in\e$. Thus $R(\nabla)$ is an {\it
  $\a$-linear} map from $\e$ to $\Omega^2 \ca
\displaystyle\mathop{\otimes}_{\a} \e$, i.e.,
a {\it tensor}. It uniquely extends to an $\a$-linear map from
$\oc \ce$ to $\oc \ce$.

Elements $a\in\a$ are called positive (or non-negative), $a\geq 0$, if
they are of the form
\[
a\;=\;\sum_i\;b_i^*\,b_i\;,\quad b_i\,\in\,\a .
\]
The module $\e$ is called {\it hermitian} if there is a {\it hermitian
  inner product}, $\langle\cdot, \cdot\rangle$ : $\e \times \e \to \a$
which is (by definition) a sesquilinear form on $\e$ with the
properties that

\begin{eqnarray}
{\bf i)}\qquad & &
\langle a s_1, b s_2\rangle\;=\;a\langle s_1,s_2\rangle b^*,\quad a, b
\in \a,\; s_1s_2\in\e\phantom{ZeichnungZeichnung}\nonumber\\
{\bf ii)}\qquad & &
\langle s,s\rangle\;\geq\;0\;,\quad {\rm for \ all}\quad s\,\in\,\e
\phantom{ZeichnungZeichnung} \\
{\bf iii)} \qquad & &
{\rm the\  map}\enskip s\;\mapsto \;\langle s,\cdot\rangle\enskip {\rm
  from}\enskip \e\enskip {\rm to\  the\  space,}\enskip
\e^*,\enskip {\rm of}\enskip\a{\rm-antilinear}\phantom{Zeichnung}
\nonumber \\
& & {\rm functionals\  on}\enskip\e\enskip {\rm
  is\   an\  isomorphism \ of \ left}\enskip\a{\rm-modules}\nonumber .
\end{eqnarray}
A hermitian inner product on $\e$ extends uniquely to a sesquilinear
form
\[
\langle \cdot,\cdot\rangle\;:\;\oc
\ce \times \oc\ce \longrightarrow \oc\ca
\]
on $\oc \ce$ with the property that
\begin{equation}
\langle \alpha \sigma_1,\beta \sigma_2\rangle\;=\;\alpha\,\langle
\sigma_1,\sigma_2\rangle\,\beta^*
\end{equation}
for all $\alpha,\beta$ in $\oc \ca$, $\sigma_1,\sigma_2\in\oc \ce$. \
A connection $\nabla$ on $\e$ is called {\it unitary} (or hermitian)
if, for all $s_1, s_2$ in $\e$,
\begin{equation}
d \,\langle s_1,s_2\rangle\;=\;\langle
\nabla\,s_1,s_2\rangle\;-\;\langle s_1,\,\nabla\,s_2\rangle .
\end{equation}
For homogeneous $\sigma_1$ and $\sigma_2$ in $\oc \ce$, one then has
that
\begin{equation}
d \,\langle \sigma_1,\sigma_2\rangle\;=\;\langle
\,\nabla\,\sigma_1,\sigma_2
\rangle\;-\;(-1)^{\deg\,\sigma_1\,+\,\deg\,\sigma_2}\;\langle\,
\sigma_1,\nabla\,\sigma_2\rangle .
\end{equation}

At this point, it should be noted that the spaces $\oc \ca$ and $\oc
\ce$ are ``monstrous'', and one cannot develop an interesting
non-commutative {\it differential} geometry without introducing
further structure. What we are looking for is a notion of {\it
  Lipshitz-} or {\it differentiable structure} on the non-commutative
space described by a unital $^*$-algebra $\a$. (Recall that the
definition of classical topological manifolds automatically entails
that manifolds are Lipshitz. In order to have a notion of
non-commutative manifolds, we thus need to introduce a notion of
Lipshitz structure.) Such a notion is obtained by considering a
$K$-{\it cycle} for $\a$, [1,2]: A $K$-cycle for $\a$ is given by the
following data:
\begin{description}
\item[i)] a separable Hilbert space ${\cal H}$;
\item[ii)] a (faithful) $^*$-representation $\pi$ of $\a$ by bounded
  operators on ${\cal H}$;
\item[iii)] a self-adjoint operator $D$ on ${\cal H}$, with the
  properties that $[D,\pi (a)]$ is a bounded operator on ${\cal H}$,
  for all $a\in\a$, and $e^{-\varepsilon D^2}$ is trace-class, for
  arbitrary $\varepsilon > 0$.
\end{description}

\medskip

\noindent\ub{Remark}: \ The trace-class property of $e^{-\varepsilon
  D^2}$, $\varepsilon > 0$, expresses the idea that the
non-commutative space described by $\a$ is compact. If this space is a
``continuum'' then $\a$ will be infinite dimensional. If $\pi$ is
faithful ${\cal H}$ must then be infinite dimensional, too. It then
follows that $D$ is unbounded. Property {\bf iii)} fixes a ``Lipshitz
structure'' on the non-commutative space described by $\a$.

\medskip

Henceforth we call $D$ the {\it Dirac operator}, following the
nomenclature used in the classical case, where $\a = \a_M$. If $\pi$ is
faithful we shall write $a$, instead of $\pi (a)$, for the operators
on ${\cal H}$ corresponding to $a\in\a$. A $K$-cycle $({\cal
  H},\pi,D)$ is called {\it even} if there is a unitary involution
$\Gamma$ on ${\cal H}$ $(\Gamma =\Gamma^*=\Gamma^{-1})$ with the
property that $\Gamma a = a\Gamma$, for all $a\in\a$, and $\Gamma D =
- D\Gamma$, (i.e. $D$ is odd). Physicists denote $\Gamma$ by $(-1)^F$,
$F$ =``fermion number''. Otherwise, $({\cal H},\pi,D)$ is called {\it
  odd}. An odd $K$-cycle $({\cal H},\pi,D)$ determines an even
$K$-cycle, $(\tilde{{\cal H}}, \tilde \pi, \tilde D)$, by setting
$\tilde{\cal H} = {\cal H} \oplus {\cal H}$, $\tilde \pi = \pi \oplus
\pi$,
\begin{equation}
\tilde D \;=\;{0\enskip D \choose D\enskip 0}\quad , \quad \Gamma
\;=\; {1\quad 0 \choose 0  -1} .
\end{equation}
A $K$-cycle $({\cal H},\pi,D)$ for $\a$ permits us to define a
$^*$-representation, $\pi$, of $\oc\ca$ on ${\cal H}$:
\begin{equation}
\pi\,(a^0da^1\cdots da^n)\;=\;a^0 [D,a^1]\cdots [D,a^n].
\end{equation}
Using the Leibniz rule, one shows that the graded subcomplex $\ker \pi
+ d\ker \pi$ of $\oc\ca$ is a two-sided ideal in $\oc\ca$,
[2,11]. Thus the quotient
\begin{equation}
\oc_D\ca\;:=\;\oc\ca \diagup (\ker \pi\,+\,\ker \pi )
\end{equation}
is a graded differential algebra. We define
\begin{eqnarray}
\oc_\pi\ca &:=& \pi(\oc\ca) \nonumber \\
{\rm Aux} &:=& \pi (d \ker \pi) \\
\Omega_{\pi,D}^n\ca &:=& \Omega_\pi^n \ca \diagup {\rm Aux}^n\nonumber \\
\oc_{\pi,D} \ca &:=& \bigoplus_{n=0}^\infty \Omega_{\pi,D}^n \ca\nonumber
\end{eqnarray}
where ${\rm Aux}^n$ is the image of all elements of $d\ker  \pi$ of
degree $n$. One calls ${\rm Aux}$ the space of ``auxiliary fields'',
[12]. Note that elements of $\oc_{\pi,D} \ca$ are equivalence classes
of bounded operators on ${\cal H}$ modulo operators in Aux. Whenever
there is no danger of confusion we omit reference to the
representation $\pi$. We note that $\oc_\pi \ca$ and Aux are left and
right modules over $\a$. Therefore $\oc_{\pi,D} \ca$ is a left and
right $\a$-module.

Next, we introduce a notion of {\it integration} on non-commutative
spaces. Given $\a$ and a $K$-cycle $({\cal H},\pi,D)$ for $\a$, we
define the integral of a form $\alpha \in \oc\ca$ by setting
\begin{equation}
  \cint \; \alpha \;:=\; \displaystyle\mathop{{\rm
      Lim}_\omega}_{\varepsilon\to 0}\; \frac{tr\,\bigl(\pi(\alpha)
    \exp (-\varepsilon\,D^2)\bigr)}{tr\,\bigl( \exp
    (-\varepsilon\,D^2)\bigr)}
\end{equation}
where \ Lim$_\omega$ denotes a limit defined in terms of some kind of
Cesaro mean, see [2]. It must then be checked that
$\cint (\cdot)$ is {\it cyclic}, i.e.,
\begin{equation}
\cint \; \alpha \beta \;=\;\cint \; \beta \alpha .
\end{equation}
Formally, this is obvious, and, in the examples we shall consider,
eq.~(2.18) will be apparent. For general results we refer the reader to
[2]. It is clear that $\cint (\cdot)$ defines a
non-negative linear functional on $\oc\ca$. Thus it determines a
positive semi-definite inner product on $\oc\ca$:
\begin{equation}
(\alpha,\beta)\;:=\;\cint\;\alpha\beta^*,\qquad \alpha,
\beta\,\in\,\oc\ca .
\end{equation}
The closure of $\oc\ca$, modulo the kernel of $(\cdot,\cdot)$, in the
norm determined by $(\cdot,\cdot)$ is a Hilbert space denoted by
$\tilde{{\cal H}} \equiv L^2 (\oc\ca)$, the Hilbert space of ``square
integrable forms''. Clearly, there is a $^*$-representation,
$\tilde\pi$, of $\oc\ca$, in particular of $\a$, on $\tilde{{\cal
    H}}$. The Hilbert space $\tilde{{\cal H}}$ has a
filtration into subspaces,
\begin{equation}
\tilde{{\cal H}}_0\; \subseteq\; \tilde{{\cal H}}_1\;\subseteq\; \cdots
\; \subseteq\;\tilde{{\cal H}}_n\;\subseteq\;\cdots\;
\subseteq\;\tilde{{\cal H}}
\end{equation}
where $\tilde{{\cal H}}_n$ is the subspace of $\tilde{{\cal H}}$
obtained by taking the closure of $\displaystyle\mathop{\cup}_{k=0}^n
\Omega^k \ca$, modulo the kernel of $(\cdot,\cdot)$, in the norm
determined by $(\cdot,\cdot)$. We denote by $\bar{\a}$ the weak
closure of $\tilde\pi(\a)$ on $\tilde{{\cal H}}_0$.
Let $P_D^{(n)}$ denote the orthogonal
projection onto $\tilde{{\cal H}}_n$, and $P_{d\ker_n}$ the orthogonal
projection onto the image of $d\ker\pi|_{\Omega^n\ca}$ in
$\tilde{{\cal H}}_{n+1}$. Given an element $\alpha \in \Omega^n \ca$, we
may define a canonical representative $\alpha^\bot$ in the image
of the equivalence class $[\alpha]\in\Omega_D^n \ca$ in $\tilde{{\cal H}}_n$
by
\begin{equation}
\alpha^\bot\;=\;(1\,-\,P_{d\ker_{n-1}})\;\alpha\,\in\,
\tilde{{\cal H}}_n .
\end{equation}
For $\alpha$ and $\beta$ in $\Omega_D^n\ca$, we set
\begin{equation}
(\alpha,\beta)\;:=\; (\alpha^\bot, \beta^\bot).
\end{equation}
We define
\begin{eqnarray}
\toc\ca &:=& \tilde\pi \bigl(\oc\ca\bigr)\nonumber\\
\tilde\Omega_D^n \ca &:=& \tilde\pi\bigl( \Omega^n \ca\bigr) \diagup
\tilde\pi (d\,J_{n-1})
\end{eqnarray}
where $J_n$ is the intersection of the kernel of $(\cdot,\cdot)$ with
$\Omega^n\ca$. By construction, $\toc \ca$ and $\toc_D
\ca$ are left and right $\a$-modules.

If $\a$ is supposed to describe something like a ``finite-dimensional,
compact, non-commutative manifold'' we must assume that
\begin{equation}
 \tilde\Omega_D^1 \ca \ {\rm is \ a \ \ub{finitely \
    generated},\ \ub{projective \ left}} \ \a{\rm -\ub{module}}.
\end{equation}
We then call $\tilde\Omega_D^1 \ca$ (the space of sections of) the
{\it cotangent bundle} of (the non-commut\-at\-ive manifold described by)
$\a$. One would then expect, moreover, that $\tilde\Omega_D^n \ca$ is
empty, for all sufficiently large $n$. In ``infinite-dimensional''
situations, encountered e.g. in string theory, $\tilde\Omega_D^1 \ca$
will of course not be finitely generated, anymore, and the theory
becomes rather tricky.

In order to develop an analogue of Riemannian geometry in the
non-commutative case, we should like to equip $\tilde\Omega_D^1 \ca$
with a {\it metric}, corresponding to a hermitian inner product on
$\tilde\Omega_D^1 \ca$. It has been shown in [3,9] that
$\tilde\Omega^1\ca$ -- in fact, $\toc\ca$ --{\it is} equipped
with a {\it canonical metric}, $\langle\cdot,\cdot\rangle_D$, (a
generalized hermitian inner product) uniquely determined by $({\cal
  H}, \pi, D)$:
\ For $\alpha$ and $\beta$ in $\toc \ca$, we set
\begin{equation}
\langle \alpha,\beta\rangle_D\;:=\; P_D^{(0)}
(\alpha\beta^*)\,\in\,\bar{{\cal A}} .
\end{equation}
A priori, $P_D^{(0)} (\alpha\beta^*)$ is just a vector in the subspace
$\tilde{{\cal H}}_0$. However, it turns out that every vector in
$\tilde{{\cal H}}_0$ uniquely corresponds to an operator on
$\tilde{{\cal H}}_0$ affiliated with the von Neumann algebra
$\bar{{\a}}$; see [3,9]. This may sound familiar from conformal field
theory. The proof follows from the cyclicity of
$\cint (\cdot)$. One then easily verifies that
$\langle\cdot,\cdot\rangle_D$ satisfies (2.9) (with the possible
exception of ({\bf iii)}) and is non-degenerate on $\toc\ca$;
see [3].
Thus, it defines a ``generalized'' hermitian inner product, or {\it
  metric} on $\toc\ca$. If $J_0=\{ 0\}$, it follows, as a
special case, that the cotangent bundle, $\tilde\Omega_D^1 \ca$,
carries a canonical metric. In the sequel, we shall assume that
property {\bf iii)} of (2.9) is satisfied, i.e., for each
$\varphi\in\tilde\Omega_D^1\ca^*$ there is a
$\tilde\varphi\in\tilde\Omega_D^1\ca$ such that
$\varphi(\omega)=\langle\omega,\tilde\varphi\rangle_D$ holds for all
$\omega\in\tilde\Omega_D^1\ca$.

By (2.24), $\tilde\Omega_D^1\ca$ is a vector bundle over $\a$. We may
thus proceed to study connections on the cotangent bundle
$\tilde\Omega_D^1\ca$. For each element $[\tilde\pi(\alpha)]
\,\in\,\tilde\Omega_D^n\ca$ we define
\begin{equation}
d[\tilde\pi(\alpha)]\;:=\;[\tilde\pi(d\alpha)]\,\in\,
\tilde\Omega_D^{n+1}\ca
\end{equation}
and it follows that $\toc_D\ca =
\displaystyle\mathop{\oplus}_{n=0}^\infty \tilde\Omega_D^n \ca$ is a
differential algebra. Definition (2.26) allows us to carry over the
tools developed in eqs.~(2.3) through (2.12) to the present context,
setting ${\cal E} =\tilde\Omega_D^1\ca$ and replacing $\oc\ca$ by
$\toc_D \ca$, and $\oc ({\cal E})$ by
\begin{equation}
\toc_D\;:=\;\toc_D \ca
\displaystyle\mathop{\otimes}_{{\a}} \;\tilde\Omega_D^1
\ca .
\end{equation}
A connection, $\nabla$, on $\tilde\Omega_D^1\ca$ is then a ${\Bbb
  C}$-linear map from $\tilde\Omega_D^1 \ca$ to $\tilde\Omega_D^1 \ca
\displaystyle\mathop{\otimes}_{{\a}}\tilde\Omega_D^1 \ca$ satisfying
\begin{equation}
\nabla (a\tilde\alpha)\;=\;da \otimes
\tilde\alpha\,+\,a\,\nabla\tilde\alpha
\end{equation}
for $a\in\a$, $\tilde\alpha\in\tilde\Omega_D^1\ca$. As above we can
extend the definition of $\nabla$ to the space $\toc_D$. The
Riemann curvature of $\nabla$ is then defined by
\begin{equation}
R(\nabla)\;=\;-\,\nabla^2 .
\end{equation}
We shall say that $\nabla$ is {\it unitary} if the formal equation
\begin{equation}
d \langle\tilde\alpha,\tilde\beta\rangle_D\;=\;
\langle\nabla\,\tilde\alpha,\tilde\beta\rangle_D\;-\;
\langle\tilde\alpha,\nabla\,\tilde\beta\rangle_D
\end{equation}
is satisfied, for all $\tilde\alpha,\tilde\beta\,\in\,\tilde\Omega_D^1
\ca$, in a sense to be made precise in more specific contexts. (The
problem in interpreting eq.~(2.30) is that
$\langle\tilde\alpha,\tilde\beta\rangle_D$ need not, in general, be an
element of the algebra $\a$ -- it belongs to the weak closure of $\a$
on $\tilde{{\cal H}}_0$ -- so the definition of the differential of
$\langle\tilde\alpha,\tilde\beta\rangle_D$ is not, a priori, clear.)

Since, by (2.24), $\tilde\Omega_D^1\ca$ is a finitely generated
projective left $\a$-module, there are generators $\{ E^A\} \subset
\tilde\Omega_D^1 \ca$, and $\{ \varepsilon_A\} \subset
\tilde\Omega_D^1 \ca^*$, $A=1,\cdots,n$, such that
\begin{equation}
\tilde\alpha\;=\;\sum_{i=1}^n\;\varepsilon_A\, (\tilde\alpha)\,E^A
\end{equation}
for any $\tilde\alpha\,\in\,\tilde\Omega_D^1\ca$; see [13]. The
Riemann curvature $R(\nabla)$ is an $\a$-linear map from
$\tilde\Omega_D^1\ca$ to the left $\a$-module $\tilde\Omega_D^2 \ca
\displaystyle\mathop{\otimes}_{{\a}} \tilde\Omega_D^1 \ca$ and one can thus
write
$R(\nabla)$ as follows (see [13]):
\begin{equation}
R(\nabla)\;=\;\sum_{A,B}\;\varepsilon_A \displaystyle\mathop{\otimes}_{{\a}}
R^A_{\;B}\displaystyle\mathop{\otimes}_{{\a}} E^B
\end{equation}
where $R_{\;B}^A\,\in\,\tilde\Omega_D^2\ca$. For an arbitrary element
$\tilde\alpha\,\in\,\tilde\Omega_D^1\ca$, $R(\nabla)\tilde\alpha$ is given
by
\begin{equation}
R(\nabla)\tilde\alpha\;=\;\sum_{A,B}\;\varepsilon_A
(\tilde\alpha)\;R_{\;B}^A \displaystyle\mathop{\otimes}_{{\a}} E^B,
\end{equation}
i.e., it belongs to $\tilde\Omega_D^2 \ca \displaystyle\mathop{\otimes}_{{\a}}
\tilde\Omega_D^1 \ca$ and is $\a$-linear in $\tilde\alpha$. It follows
from  properties (2.9), of the metric
$\langle\cdot,\cdot\rangle_D$ that the map
\[
\tilde\Omega_D^1\ca \rightarrow
\tilde\Omega_D^1\ca^*\;,\quad\tilde\alpha\mapsto
\langle\cdot,\tilde\alpha^*\rangle_D,
\]
is an isomorphism of right $\a$-modules. Thus, for each
$A=1,\cdots,n$, we can define an element
$\tilde\varepsilon_A\,\in\,\tilde\Omega_D^1\ca$ by
\begin{equation}
\varepsilon_A(\tilde\alpha)\;=\;\langle\tilde\alpha,
\tilde\varepsilon_A^*\rangle_D\;,
\quad {\rm for \ all}\enskip \alpha\,\in\,\tilde\Omega_D^1 \ca .
\end{equation}
The {\it Ricci tensor} associated with the connection $\nabla$ can
then be defined {\it invariantly} by
\begin{equation}
Ric\, (\nabla)\;=\;\sum_{A,B}\;(P_D^{(1)}-P_D^{(0)})\,
(\tilde\varepsilon_A\,R_{\;B}^{A\,\bot})\;
\displaystyle\mathop{\otimes}_{\a}\; E^B\,\in\,\tilde{{\cal
    H}}_1\;\displaystyle\mathop{\otimes}_{\a}\,\tilde\Omega_D^1 \ca
\end{equation}
where $R_{\;B}^{A\,\bot} = (1-P_{d\,J_1})\,R_{\;B}^A$, and $P_{dJ_1}$
is the orthogonal projection onto the closure of $dJ_1$. Notice that
$\a$ acts on $\tilde{{\cal H}}$ from the right due to the cyclicity of
$\cint \; (\cdot)$. The {\it scalar curvature}, $r (\nabla)$, of the
connection $\nabla$ can now be defined by
\begin{equation}
r \,(\nabla)\;=\; \sum_{A,B}\;P_D^{(0)}\;\bigl( (P_D^{(1)}-P_D^{(0)})
(\tilde\varepsilon_A\, R_{\;B}^{A\,\bot})\;E^B\bigr).
\end{equation}
These definitions are discussed in [14]; (see also [9,11] for a
preliminary account.)

Following [3], we define the {\it torsion}, $T(\nabla)$, of the
connection $\nabla$ by
\begin{equation}
T (\nabla)\;=\;d\,-\,m \circ \nabla
\end{equation}
where $m$ is multiplication of forms. One verifies without difficulty
that $T(\nabla)$ is an $\a$-linear map from $\toc_D$ to
$\toc_D$, (i.e., $T(\nabla)$ is a tensor).
A connection $\nabla$ is called a {\it Levi-Civita connection} if
$\nabla$ is unitary and $T(\nabla)=0$. In contrast to the classical
case, there are ``non-commutative Riemannian spaces'' $(\a,{\cal
  H},\pi,D)$ which do not admit any Levi-Civita connection and ones
that admit many.

In our calculations in Sect.~3, we shall make use of the
non-commutative analogue of the {\it Cartan structure equations} which
were found in [3]. However, since the cotangent bundle
$\tilde\Omega_D^1 \ca$ is, in general, not a free left $\a$-module, we
need a slight generalization of these equations. Here, we only state
results (for detailed proofs, see [14]).

The components $\Omega_{\;B}^A = \Omega_{\;B}^A
(\nabla)\,\in\,\tilde\Omega_D^1\ca$ of a connection $\nabla$ on
$\tilde\Omega_D^1\ca$ are defined by
\begin{equation}
\nabla E^A\;=\;-\,\Omega_{\;B}^A \,\otimes\,E^B
\end{equation}
where we use the summation convention and drop the subscript $\a$
on the tensor product symbol. Since the generators $\{ E^A\}$ of
$\tilde\Omega_D^1\ca$ are, in general, not linearly independent over
$\a$,
the coefficients $\Omega_{\;B}^A$ cannot be chosen arbitrarily, and
are not unique in general. However, for any matrix
$\tilde\Omega_{\;B}^A\,\in\,\tilde\Omega_D^1\ca$, the coefficients
\begin{equation}
\Omega_{\;B}^A\;=\;\varepsilon_C
(E^A)\,\tilde\Omega_{\;D}^C\;\varepsilon_B (E^D)\;-\;d\,\varepsilon_B
(E^A)
\end{equation}
define a connection on $\tilde\Omega_D^1 \ca$, and every connection is
of this form. The components of $T(\nabla)$ and $R(\nabla)$ are
defined by
\begin{equation}
T(\nabla)\,E^A\;=\;T^A\,\in\,\tilde\Omega_D^2\,\ca
\end{equation}
and
\begin{equation}
R(\nabla)\,E^A\;=\;R_{\;B}^A\,\otimes\,E^B,
\end{equation}
where $R_{\;B}^A\,\in\,\tilde\Omega_D^2 \ca$. Notice that the
components, $R_{\;B}^A$, of the curvature are not uniquely defined, in
general. Combining (2.37) with (2.40) we find that
\begin{equation}
  T^A\;=\;d
  E^A\;+\;\Omega_{\;B}^A\,E^B\;=\;\varepsilon_B(E^A)\,dE^B\;+\;
  \stackrel{\approx}{\Omega}\,\!\!_{\;B}^A\,E^B
\end{equation}
where $\stackrel{\approx}{\Omega}\,\!\!_{\;B}^A = \varepsilon_C (E^A)\,
\tilde\Omega_D^C \, \varepsilon_B (E^D)$. From (2.28), (2.29) and (2.41)
we obtain that
\begin{eqnarray}
R_{\;B}^A &=& d\,\Omega_{\;B}^A\,+\,\Omega_{\;C}^A\;\Omega_{\;B}^C
\nonumber\\
&=& d\,\stackrel{\approx}{\Omega}\,\!\!_{\;B}^A\,+\,
\stackrel{\approx}{\Omega}\,\!\!_{\;C}^A \;
\stackrel{\approx}{\Omega}\,\!\!_{\;B}^C\;+\; d\varepsilon_C (E^A)\,
d\varepsilon_B(E^C) .
\end{eqnarray}
Eqs.~(2.42) and (2.43) are the non-commutative Cartan equations.

For a Riemannian manifold, the Levi-Civita connection is invariant
under all one-parameter groups of isometries. Since, in the
non-commutative setting, there are often a lot of Levi-Civita connections,
it is useful to look at connections which are also invariant under
isometries. A {\it one-parameter group of isometries} of the
``non-commutative Riemannian space'' $(\a, {\cal H},\pi,D)$ is a
one-parameter group of unitaries $ U(t)$ on ${\cal H}$
such that, for all $t\in{\Bbb R}$,
\begin{eqnarray}
U(t)\,\a\,U(t)^*   &=& \a \nonumber\\
\, [ D, \,U(t) ]      &=& 0\;.
\end{eqnarray}
A connection $\nabla$ is said to be invariant under $U(t)$ if it
satisfies
\begin{equation}
\nabla\bigl(\,U(t)\,\tilde\alpha\,U(t)^*\bigr)\;=\;\bigl(
U(t)\,\otimes\,U(t)\bigr)\;\nabla\tilde\alpha\;\bigl(U(t)^*\,\otimes\,
U(t)^* \bigr),
\end{equation}
for any $\tilde\alpha \in \tilde\Omega_D^1 ({\cal A})$.

\medskip

\noindent This completes our review of non-commutative Riemannian geometry.

\vspace{1cm}

\section{The non-commutative Riemannian geometry\hfill\break
of the standard model}

The construction of the standard model in non-commutative geometry
[12,1,2,5] requires an appropriate choice of a non-commutative
Riemannian space $(\a,{\cal H},\pi,D)$, as defined in the last
section. The algebra $\a$ defining the non-commutative space
underlying the standard model is chosen to be
\begin{equation}
\a\;=\;(\a_1\,\oplus\,\a_2)\;\otimes\;C^\infty\,(M_4),
\end{equation}
where $M_4$ is a smooth, compact, four-dimensional Riemannian spin
manifold, $\a_1 = {\Bbb M}_2 ({\Bbb C})$ is the algebra of complex 2$\times$2
matrices, and $\a_2={\Bbb C}$. (We shall only consider the leptonic
and Higgs sector of the standard model and omit quarks and
gluons. They could be included in our analysis, but merely complicate
our formulas.) Elements, $a$, of $\a$ are written as
\begin{equation}
a\;=\;{a_1\enskip 0 \choose 0\enskip a_2}\enskip ,
\end{equation}
where $a_i$ is a $C^\infty$-function on $M_4$ with values in $\a_i$,
$i=1,2$.

The Hilbert space is defined to be
\begin{equation}
{\cal H}\;=\;L^2\,(S_1, dv_1)\;\oplus\;L^2\,(S_2,dv_2),
\end{equation}
where $S_i = S_0 \otimes V_i$, $S_0$ is the usual bundle of Dirac
spinors on $M_4$, and $V_i$ is a representation space for $\a_i$,
$i=1,2$, with $V_1={\Bbb C}^2$ and $V_2={\Bbb C}$, and $dv_i$ is the
volume form corresponding to a Riemannian metric $g_i$ on $M_4$, with
$i=1,2$. Thus $\a$ acts on sections of $S_1\oplus S_2$ by left
multiplication, and ${\cal H}$ is a left $\a$-module of
square-integrable $V_1\oplus V_2$-valued Dirac spinors on $M_4$. The
representation $\pi$ of $\a$ is given by
\begin{equation}
\pi\;=\;\pi_1\;\oplus\;\pi_2 ,
\end{equation}
where $\pi_i$ is the representation of $\a_i \otimes C^\infty (M_4)$
on $L^2\,(S_i,dv_i)$ given by left-multiplication of sections of $S_i$
by elements of $\a_i \otimes C^\infty (M_4)$.

The Dirac operator $D$ is given by
\begin{equation}
D\;=\; \left(\begin{array}{cc}
\nabla\!\!\!\!/\,_1 \otimes 1_2 \otimes 1_3 \enskip & \gamma^5 \otimes M_{12}
\otimes k \\
\gamma^5\otimes M_{12}^*\otimes k^* & \nabla\!\!\!\!/\,_2\otimes 1_3
\end{array} \right)\; ,
\end{equation}
where $\nabla\!\!\!\!/_i$ is the covariant Dirac operator on $L^2
(S_0,dv_i)$; in a coordinate chart, $U$, of $M_4$
\begin{equation}
\nabla\!\!\!\!/\,_i\;=\;e_{ia}^\mu\,\gamma^a\,(\partial_\mu +
i\omega_{i\mu} ),
\end{equation}
where $\{ e_{i\,a}^\mu\}$ is a vierbein, i.e., an orthonormal basis of
sections of the tangent bundle $TU$, so that
\begin{eqnarray}
e_{i\,a}^\mu\;g_{i\mu\nu}\;e_{i\,b}^\nu &=& \delta_{ab},\nonumber\\
e_{i\,a}^\mu\;\delta^{ab}\;e_{i\,b}^\nu &=& g_i^{\mu\nu};\nonumber
\end{eqnarray}
$\omega_{i\mu}=\frac 1 4\;\omega_{\mu a b} (e_i) [\gamma^a,\gamma^b]$
is the corresponding spin connection, with $\omega_{\mu a b} (e_i)$ a
solution of the Cartan structure equation
\[
T_{ia}\;=\;de_{ia} + \sum_b\;\omega_{ab}\;(e_i)\;e_b\;=\;0,
\]
for a unitary connection on $TU$, $i=1,2$; and $\{\gamma^a\}_{a=1}^4$
are the anti-hermitian Euclidian Dirac matrices, with \ $\{ \gamma^a,
\gamma^b\} = \gamma^a\gamma^b+\gamma^b\gamma^a= - 2\delta^{ab}$,
$\gamma^5=\gamma^1\gamma^2\gamma^3\gamma^4$; (we note that
$(\gamma^5)^*=\gamma^5$). Furthermore, $k$ is a 3$\times$3 family
mixing matrix, and
\begin{equation}
M_{12}\;=\;{\alpha\, (x) \choose \beta\,(x)}\;,\quad M_{21}\;:=\;M_{12}^*
,
\end{equation}
where $\alpha$ and $\beta$ are smooth, complex-valued function on
$M_4$; ($M_{12}$ is called a ``doublet'').

Next, we study the algebra, $\oc_D\ca $, of differential forms for
$\a$. A 1-form $\rho = \sum_i a_i\, db_i\,\in\,\Omega^1\ca$ is
represented on ${\cal H}$ as the operator
\begin{equation}
\pi (\rho)\;=\;\sum_i\;a_i\, [D, b_i]\;=\; \left(
\begin{array}{cc}
\gamma^a\,A_{1a}\enskip & \gamma_5\,k\,\phi_{12} \\
\gamma_5\,k^*\,\phi_{21}\enskip & \gamma^a\,A_{2a}
\end{array} \right)\enskip ,
\end{equation}
where
\begin{eqnarray}
A_{1a}&=&e_{1a}^\mu\;\sum_i\;a_{1i}\;\partial_\mu\;b_{1i},\nonumber\\
A_{2a}&=&e_{2a}^\mu\;\sum_i\;a_{2i}\;\partial_\mu\;b_{2i},\nonumber\\
\phi_{12}&=&\sum_i\;a_{1i}\;M_{12}\;b_{2i}-M_{12},\\
\phi_{21}&=&\sum_i\;a_{2i}\;M_{21}\;b_{1i}-M_{21},\nonumber
\end{eqnarray}
where we have assumed, without loss of generality, that
\begin{equation}
\sum a_{1i}\;b_{1i}\;=\;1,\;\sum a_{2i}\;b_{2i}\;=\;1 \; .
\end{equation}

We also need to understand the space, $\Omega_D^2 \ca$, of
2-forms. Since $\pi$ is faithful, $\Omega_D^2 \ca$ is isomorphic to
$\pi \bigl(\Omega^2\,\ca\,\bigr)\diagup {\rm Aux}^2$,  where
\begin{equation}
{\rm Aux}^2\;=\;\bigl\{ \sum_i
\,[D,a_i][D,b_i]\;:\;\sum_i\;a_i\,[D,b_i]\;=\;0\bigr\},
\end{equation}
see (2.16). For $\rho = \sum_i\,a_i\,db_i\,\in\,\ker \pi$, $\pi
(d\rho) = \sum_i\,[D,a_i][D,b_i]$ can be evaluated by using
eqs.~(3.2), (3.5) and (3.9). After some algebra one finds that
$\pi(d\rho)$, written as a 2$\times$2 matrix, has the following
entries:

\begin{eqnarray}
\pi\,(d\rho)_{11} &=& \partial\!\!\!/\,A_1 -
\sum\,a_{1i}\,\partial\!\!\!/_1^2\,b_{1i}\nonumber \\
&+& k\,k^*\,\bigl( M_{12}
(\phi_{21}+M_{21})+(\phi_{12}+M_{12})\,M_{21}-2
M_{12}\,M_{21}\nonumber\\
&-& \sum\,a_{1i}\, [M_{12}\,M_{21},b_{1i}]\bigr),\nonumber\\
\pi\,(d\rho)_{22} &=& \partial\!\!\!/\,A_2 -
\sum\,a_{21}\,\partial\!\!\!/_2^2\,b_{2i}\nonumber\\
&+&
k^*k\,\bigl(M_{21}\,(\phi_{12}+M_{12})+(\phi_{21}+M_{21})\,M_{12}-2
M_{21}\,M_{12}\nonumber \\
&-& \sum\,a_{2i}\,[M_{21}\,M_{12},b_{2i}]\bigr), \\
\pi\,(d\rho)_{12} &=& \gamma^5k\, \bigr[-
A_1\,M_{12}+M_{12}\,A_2-\partial\!\!\!/_1\,(\phi_{12}+M_{12})\nonumber\\
&+& \sum\,a_{1i}\,M_{12}\,(\partial\!\!\!/_1 - \partial\!\!\!/_2)\,
b_{2i}+\sum\,
a_{1i}\,(\partial\!\!\!/_1\,M_{12})\,b_{2i}\bigr],\nonumber\\
\pi\,(d\rho)_{21} &=& \gamma^5 k^*\,\bigl[ -A_1\,M_{21}+M_{21}\,A_1 -
\partial\!\!\!/_2\,(\phi_{21}+M_{21})\nonumber\\
&+& \sum\,a_{2i}\,M_{21}\,(\partial\!\!\!/_2\,-
\partial\!\!\!/_1\,)\,b_{1i} + \sum\,a_{2i}\,
(\partial\!\!\!/_2\,M_{21})\,b_{1i}\bigr],\nonumber
\end{eqnarray}
where $\partial\!\!\!/_i = e_{i\,a}^\mu\;\gamma^a\;\partial_\mu$,
$i=1,2$.

Assuming that $\rho\,\in\,\ker \pi$, i.e., $\pi (\rho)=0$, eqs.~(3.12)
reduce to
\begin{eqnarray}
\pi\,(d\rho)_{11} &=& - \sum\,a_{1i}\,\partial\!\!\!/_1^2\,b_{1i} -
k\,k^*\,\sum\,a_{1i}\, [M_{12}\,M_{21},b_{1i}],\nonumber\\
\pi\,(d\rho)_{22} &=& - \sum\,a_{2i}\,\partial\!\!\!/_2^2\,
b_{2i}\\
\pi\,(d\rho)_{12} &=& \gamma^5 k\,\bigl[
\sum\,a_{1i}\,(\partial\!\!\!/_1\,M_{12})\,b_{2i}-\partial\!\!\!/_1\,
M_{12} + \sum\,a_{1i}\,M_{12}\,(\partial\!\!\!/_1 -
\partial\!\!\!/_2)\,b_{2i}\bigr],\nonumber\\
\pi\,(d\rho)_{21} &=& \gamma^5 k^*\,\bigl[
\sum\,a_{2i}\,(\partial\!\!\!/_2\, M_{21})\,b_{1i}-\partial\!\!\!/_2\,
M_{2i} + \sum\,a_{2i}\,M_{21}\,(\partial\!\!\!/_2 -
\partial\!\!\!/_1)\,b_{1i}\bigr]\;.\nonumber
\end{eqnarray}
Thus $\pi(d\rho)|_{\rho=0}$ is an operator of the form
\begin{equation}
\left( \begin{array}{cc}
X_1+k\,k^*\,Y_1\quad & \gamma^5\,k\,\gamma^a\,X_{1a}\\
\gamma^5 k^*\gamma^a X_{2a} & X_2
\end{array} \right)
\end{equation}
where $Y_1$ is an arbitrary function on $M_4$. (We have simplified our
notations by omitting writing the identity element of the Clifford
algebra and the tensor product symbols.) The structure of the space of
auxiliary 2-forms, Aux$_2$, depends on the properties of \ $e_{1a}^\mu
- e_{2a}^\mu$ and $\partial\!\!\!/\,M_{12}$. There are three
possibilities, namely:
\begin{description}
\item[a)] $X_{1a}$ and $X_{2a}$ are arbitrary functions. Then, the
canonical representative, $\omega^\bot$, of a 2-form $[\omega] \in
\Omega_D^2 \ca$ has vanishing off-diagonal elements and the Higgs
field is not dynamical.
\item[b)] $X_{1a}$ and $X_{2a}$ are neither arbitrary nor identically
  zero. In this case, the evaluation of $\Omega_D^2 \ca$ at a point
  $p\in M_4$, $\Omega_D^2 \ca (p)$, is not everywhere isotropic and
  its structure may depend on the evaluation point.
\item[c)] $X_{1a}$ and $X_{2a}$ are identically zero, i.e. auxiliary
  2-forms are diagonal. Then the Higgs field is dynamical, $\Omega_D^2
  \ca (p)$ is everywehere isotropic and its structure doesn't
  depend on $p$. This is the case we shall consider in the sequel.
\end{description}
Next we compute the
constraints implied by the vanishing of $X_{1a}$ and $X_{2a}$. Using
(3.13) we see that the condition $\pi(d\rho)_{12}|_{\rho=0}= \pi
(d\rho)_{21}|_{\rho=0}=0$  is
equivalent to
\begin{eqnarray}
&& e_{1a}^\mu\;-\; e_{2a}^\mu\;=\; 0\nonumber\\
&&\sum_i\;a_{1i}\,(\partial\!\!\!/\,M_{12})\,b_{2i}\;-\;\partial\!\!\!/\,
M_{12}\;=\;0\\
&&\sum_i\;a_{2i}\,(\partial\!\!\!/\,M_{21})\,a_{1i}\;-\;
\partial\!\!\!/\,M_{21}\;=\;0\nonumber
\end{eqnarray}
whenever $\pi(\rho)=0$. The second equation implies that
\begin{equation}
\partial\!\!\!/\,M_{12}\;=\;\gamma^\mu\, c_\mu\,M_{12}
\end{equation}
for some functions $c_\mu$, and using equation (3.7) one easily proves
that
\begin{equation}
M_{12}\;=\;e^{-\sigma}\; {\alpha_0 \choose \beta_0}
\end{equation}
where $\sigma$ is an arbitrary complex valued function and
$\alpha_0,\beta_0$ are constant (we exclude the case
$\alpha_0=\beta_0=0$ since this would lead to diagonal 2-forms). We
can set without loss of generality
\begin{equation}
\alpha_0\;=\;0, \quad
\beta_0\;=\;1, \quad
Im\,\sigma\;=\;0 .
\end{equation}
This can be seen by considering the unitarily equivalent $K$-cycle
$({\cal H}, \a, UDU^*)$ where
\begin{equation}
U\;=\;
\left( \begin{array}{ccc}
\triangle\beta_0 & -\triangle \alpha_0 & 0 \\
\triangle\bar\alpha_0 & \triangle\bar\beta_0 & 0 \\
0 & 0 & e^{i\,Im\,\sigma}
\end{array}\right)
\end{equation}
and $\triangle = (\mid\alpha_0\mid^2 + \mid\beta_0\mid^2)^{-1/2}$. The
new Dirac operator reads
\begin{equation}
UDU^*\;=\;
\left( \begin{array}{cc}
\nabla\!\!\!\!/ & k\,\gamma^5\,e^{-Re\,\sigma}{0 \choose 1} \\
k^* \gamma^5\,e^{-Re\,\sigma}\,(0,1)\quad & \nabla\!\!\!\!/
\end{array}\right)\; +\;
\left( \begin{array}{cc}
0 & 0 \\
0 & i\,\partial\!\!\!/\, Im\,\sigma_2
\end{array} \right)
\end{equation}
and we can drop the second term since it doesn't contribute to
commutators with elements of $\a$. The result of these computations is
that the existence of a dynamical Higgs field and local isotropy imply
that the distance between the two sheets is described by a single real
scalar field, $\sigma$, and that the metrics on them must be
identical.

It is not hard to show that, in our case,
\begin{equation}
\cint \alpha\;=\; c\bigl( \int\limits_{M_4} tr_1\,(\pi
(\alpha)_{11})\,dv_1\,+\, \int\limits_{M_4}
tr_2\,(\pi(\alpha)_{22})\,dv_2\bigr)
\end{equation}
where $tr_1$ and $tr_2$ are normalized on the generation space such
that
\begin{equation}
tr_i\,(k\,k^*)\;=\;1,\quad tr_i\,(\id_3)\;=\;1,\quad i\;=\;1,2
\end{equation}
and are standard traces on the Clifford algebra and ${\Bbb M}_2 ({\Bbb
  C})$. The constant $c$ is chosen such that $\cint 1=1$. Equation
(3.21) follows from results in [2]. It follows that
\begin{equation}
(\alpha,\alpha)\;=\;\cint \alpha\alpha^*\;=\;0\qquad {\rm iff}\enskip
\pi(\alpha)\;=\;0,
\end{equation}
for all $\alpha\,\in\,\oc\ca$. This means that $(\cdot,\cdot)$ has a
trivial kernel in $\oc_\pi \ca$ (see (2.16)), and hence the
representation $\tilde\pi$, defined before eq.~(2.20), is a faithful
representation of the algebras $\oc_\pi\ca$ and $\oc_D\ca$ (the
algebra of differential forms) and, in particular of $\a$.

{}From the fact that the Clifford algebra generated by the Dirac
matrices $\gamma^1, \cdots, \gamma^4$ is finite-dimensional one
deduces that there exists some $n<\infty$ such that, in eq.~(2.20),
\begin{equation}
\tilde{{\cal H}}_0\;\subset\;\tilde{{\cal H}}_1\;\subset\cdots\subset
\; \tilde{{\cal H}}_n\;=\;\tilde{{\cal H}},
\end{equation}
and hence $\Omega_D^k \ca = \{ 0\}$, for $k>n$. From (3.8) and (3.9)
we infer that $\tilde\Omega_D^1 \ca$ is a finitely generated,
projective left $\a$-module.

Next, we proceed to determine the Levi-Civita connections, i.e., the
unitary torsionless connections. To this aim, we introduce a system of
generators of $\tilde\Omega_D^1 \ca$, $\{ E^A\}$, given by
\begin{equation}
E^a\;=\;\gamma^a\; {\id_2\enskip 0 \choose 0\quad 1} \qquad
a\;=\;1,2,3,4
\end{equation}

\begin{equation}
E^r\;=\;\gamma^5\;
\left( \begin{array}{cc}
O_2 & k\;e_r \\
-\,k^*\,e_r^\top & 0
\end{array} \right)\; ,\enskip r\;=\; 5,6
\end{equation}
where $e_5 = {1\choose 0}$ and $e_6 = {0\choose 1}$. It is then easy
to check that the elements $\{ \varepsilon_A\} \subset
\tilde\Omega_D^1 \ca^*$ given by

\begin{eqnarray}
&\varepsilon_a (\omega)\;=\;
\left( \begin{array}{cc}
e_a^\mu\,\omega_{1\mu} & 0 \\
0 & e_a^\mu\,\omega_{2\mu}
\end{array} \right) \nonumber \\
&\varepsilon_5 (\omega)\;=\;
\left( \begin{array}{ccc}
\frac{\omega_1}{2} & 0 & 0 \\
\frac{\omega_2}{2} & 0 & 0 \\
0 & 0 & -\,\tilde\omega_1
\end{array} \right) \\
&\varepsilon_6(\omega)\;=\;
\left( \begin{array}{ccc}
0 & \frac{\omega_1}{2} & 0 \\
0 & \frac{\omega_2}{2} & 0 \\
0 & 0 &-\,\tilde\omega_2
\end{array}\right)\nonumber
\end{eqnarray}
for any 1-form $\omega$ written as
\[
\omega\;=\;
\left( \begin{array}{cc}
\gamma^\mu\,\omega_{1\mu} &
\begin{array}{c}
 \omega_1 \\ \omega_2
\end{array} \\
\begin{array}{cc}
\tilde\omega_1\,& \tilde\omega_2
\end{array} &
 \gamma^\mu\,\omega_{2\mu}
\end{array} \right)
\]
with $\omega_{1\mu}$ a 2$\times$2 matrix, satisfy eq.~(2.31), i.e.,
they build a ``dual basis''.

We define the connection coefficients by
\begin{equation}
\nabla E^A\;=\;\,-\,\Omega_{\;B}^A \otimes E^B
\end{equation}
where $A,B=1,\cdots,6$. The connection coefficients being 1-forms, we
write
\begin{equation}
\Omega_{\;B}^A\;=\;
\left( \begin{array}{cc}
  \gamma^\mu\,\omega_{1\mu}\;_{\;B}^A & k\,\gamma^5\,e^{-\sigma}\;
{\omega_1\;_{\;B}^A \choose \omega_2\;_{\;B}^A} \\
k^*\,\gamma^5\,e^{-\sigma}\,(\tilde\omega_1\;_{\;B}^A,\;
\tilde\omega_2\;_{\;B}^A)
& \gamma^\mu\,\omega_{2\mu}\;_{\;B}^A
\end{array} \right)
\end{equation}
where $\omega_{1\mu}\;_{\;B}^A$ is a 2$\times$2 matrix. (In the sequel
we shall omit to specify the representation,i.e., we write $\omega$
for $\tilde\pi (\omega)$ for any form $\omega$.) Since the generators,
$\{ E^A\}$, of $\tilde\Omega_D^1 \ca$ are anti-Hermitian, they
correspond to real forms. Thus, we assume the matrix elements of the
connection coefficients to be real. Since $\tilde\Omega_D^1 \ca$ is
not a free module, the coefficients $\Omega_{\;B}^A$ are not
independent. Using eq.~(2.39) one gets a large number of
constraints. These are listed in the appendix eqs.~(A.2) -
(A.4). Then, we require the connection to be unitary, i.e., (see
eq.~(2.30))
\begin{equation}
d\langle E^A, E^B\rangle_D\;=\;-\,\Omega_{\;C}^A\,\langle\,E^C,
E^B\rangle_D\;+\;\langle E^A, E^C\rangle_D\;(\Omega_{\;C}^A)^* .
\end{equation}
The products between the generators are easily computed and one finds
\begin{eqnarray}
&& \langle E^a, E^b\rangle_D\;=\;\delta^{ab}\;,\quad a,b
\;=\;1,\cdots,4 \nonumber \\
&& \langle E^a, E^r\rangle_D\;=\;0\;,\quad\enskip a\;=\;1,\cdots,4,\; r =
5,6.\nonumber\\
&& \quad \\
&& \langle E^5, E^5\rangle_D\;=\;
\left( \begin{array}{ccc}
1 & 0 & 0 \\
0 & 0 & 0 \\
0 & 0 & 1 \end{array} \right)
\qquad \langle E^5, E^6\rangle_D\;=\;
\left( \begin{array}{ccc}
0 & 1 & 0 \\
0 & 0 & 0 \\
0 & 0 & 0 \end{array} \right) \nonumber \\
&& \langle E^6, E^5\rangle_D\;=\;
\left( \begin{array}{ccc}
0 & 0 & 0 \\
1 & 0 & 0 \\
0 & 0 & 0 \end{array}\right)
\qquad \langle E^6, E^6\rangle_D\;=\;
\left( \begin{array}{ccc}
0 & 0 & 0 \\
0 & 1 & 0 \\
0 & 0 & 1 \end{array}\right)\;. \nonumber
\end{eqnarray}

\noindent Using eqs.~(3.29) and (3.30) one gets the unitarity
conditions listed in the appendix, eqs.~(A.9) - (A.11). Next, we
compute the torsion. The components of the torsion are defined by
\begin{equation}
T^A\;=\;dE^A\;+\;\Omega_{\;B}^A\;E^B.
\end{equation}
In order to compute these components, we have to know the
differentials of the generators. These are computed as follows: we
write $E^5$ and $E^6$ as
\begin{equation}
E^r\;=\;e^\sigma\;[D,m^r]\;,\quad r\;=\;5,6
\end{equation}
where $m^r\,\in\,\a$ are given by
\begin{equation}
m^5\;=\;\left( \begin{array}{rrr}
0  & -1 &  0 \\
-1 & 0  & 0 \\
0  & 0  & 0 \end{array} \right)\; , \qquad m^6\;=\;
\left( \begin{array}{rrr}
0 & 0 & 0 \\
0 &-1 & 0 \\
0 & 0 & 0
\end{array} \right)\; .
\end{equation}
One then easily checks that
\begin{equation}
d E^r\;=\;[D,e^\sigma][D,m^r]\;=\;\partial\!\!\!/\,\sigma E^r,\enskip
r=5,6 .
\end{equation}
Notice that eq.~(3.35) gives already the canonial representative,
$(dE^r)^\bot$, of $dE^r$ since the auxiliary 2-forms are diagonal. For
completeness we give the general form of the canonical representative,
$\omega^\bot$, of a 2-form $\omega \,\in\,\tilde\Omega_D^2 \ca$,
\begin{equation}
\omega^\bot\;=\;
\left( \begin{array}{cc}
\gamma^{\mu\nu}\,\omega_{1\mu\nu} &
k\,\gamma^\mu\gamma^5\,{\omega_{1\mu} \choose \omega_{2\mu}} \\
k^*\gamma^\mu\gamma^5\,(\tilde\omega_{1\mu},\tilde\omega_{2\mu} &
\gamma^{\mu\nu} \omega_{2\mu\nu} + (k^*k-1)\,\omega
\end{array} \right)
\end{equation}
where $\omega_{1\mu\nu}$ is a 2$\times$2 matrix and
$\gamma^{\mu\nu}=\frac 1 2 \,[\gamma^\mu, \gamma^\nu]$. Notice that
the only effect of the projection $\omega \to \omega^\bot$ which
differs from the classical case is the replacement $k^*k\to k^*k-1$ in
the matrix element $\omega_{22}$. Using eqs.~(3.32), (3.35) and (3.36)
one computes the components of the torsion. The conditions of
vanishing torsion are listed in the appendix, eqs.~(A.15) and (A.16),
where it is shown that the condition $T^A=0$ makes the $\sigma$-field
non-dynamical. Thus, we shall consider unitary connections for which
the following weaker condition holds,
\begin{equation}
Tr_k\,T^A\;=\;0\;,\quad A\;=\;1,\cdots,6
\end{equation}
where $Tr_k$ denotes the trace over the generation space.

The next step is to compute connections which are invariant under
isometries of the underlying non-commutative space. Since the
classical manifold $M_4$ is not specified, we don't know if it admits
any Killing field. Thus, we look for isometries described by a
one-parameter group of unitaries, $U(t)$, with constant
coefficients. It is easy to prove that the requirements
\begin{equation}
U(t)\,\a\,U(t)^*\;\subset\;\a,\; [D, U(t)]\;=\;0
\end{equation}
imply that $U(t)$ is of the form
\begin{equation}
U(t)\;=\;\left( \begin{array}{ccc}
e^{-\,it} & 0 & 0 \\
0 & e^{it} & 0 \\
0 & 0 & e^{it} \end{array} \right)\; .
\end{equation}
The transformation properties of the generators, $\{ E^A\}$, of
$\tilde\Omega_D^1 \ca$ are then
\begin{eqnarray}
&& U(t)\,E^a\,U(t)^*\;=\;E^a\;,\;U(t)\,E^6\,U(t)^*\;=\;E^6 \nonumber\\
&& \quad\\
&& U(t)\,E^5\,U(t)^*\;=\;\gamma^5\,
\left( \begin{array}{cc}
O_2 & k\,e^{-\,2i\varphi}\,e_5 \\
-\,k^*\,e^{2i\varphi}\,e_5^\top & O_1 \end{array} \right)\;. \nonumber
\end{eqnarray}

\noindent The conditions implied by the invariance of the connection
under these isometries (see eq.~(2.45)) are listed in eqs.~(A.19)-(A.21) of the
appendix.

Finally we compute the Hilbert-Einstein action for unitary connections
which are invariant under isometries and for which $Tr_k T^A
(\nabla)=0$ holds. The components of the curvature are given by (see
eq.~(2.43))
\begin{equation}
R_{\;B}^A\;=\;d\Omega_{\;B}^A + \Omega_{\;C}^A \Omega_{\;B}^C .
\end{equation}
We write the components of the curvature as
\begin{equation}
R_{\;B}^A\;=\; \left( \begin{array}{cc}
\gamma^{\mu\nu}\,R_{\mu\nu}^{(1)}\;_{\;B}^A & k\,\gamma^\mu\,\gamma^5\,
e^{-\sigma}\,P_\mu\;_{\;B}^A \\
k^*\,\gamma^\mu\,\gamma^5\,e^{-\sigma}\,Q_\mu\;_{\;B}^A\quad &
\gamma^{\mu\nu}\,R_{\mu\nu\;B}^{(2)\,A}\,+\,(k^*k-1)\,L_{\;B}^A
\end{array} \right) \;.
\end{equation}
Explicit formulas for these quantities are given in the appendix,
eqs.~(A.22) - (A.24). The 1-forms, $\tilde\varepsilon_A$,
corresponding to the dual forms, $\varepsilon_A$, of eq.~(3.27)
through the definition (2.34) are easily shown to be
\begin{eqnarray}
&& \tilde\varepsilon_a\;=\;-\,\gamma^a\;
\left( \begin{array}{ll}
\id_2 & 0\\
0 & 1  \end{array}\right) \nonumber \\
&& \quad \\
&& \tilde\varepsilon_r\;=\;\gamma^5\;
\left( \begin{array}{cc}
O_2 & -\,k\;e_r \\
\frac 1 2 \;k^*\,e_r^\top & O \end{array} \right)\; . \nonumber
\end{eqnarray}
This allows us to compute the matrix elements of the Ricci tensor (see
eq.~2.35) and then also the scalar curvature defined in eq.~(2.36).
The explicit formula for the Ricci tensor is given in eq.~(A.25). Then
one computes the Hilbert-Einstein action
\begin{eqnarray}
&&\cint r \;=\; \int \sqrt{g}\;d^4x\;\bigl\{ \bigl( e_a^\nu\,e_b^\mu -
e_a^\mu\,e_b^\nu\bigr) \bigl(Tr\,R_{\mu\nu\;b}^{(1)\;a} +
R_{\mu\nu\;b}^{(2)\;a}\bigr)\;+\nonumber\\
&&\quad e^{-\sigma}\;\bigl[ e_a^\mu\,\bigl( \frac 1
2\;P_{1\mu\;a}^{\;5}\,+\,\frac 1 2\;
P_{2\mu\;a}^{\;6}\,-\,P_{1\mu\;5}^{\;a}\,-\,P_{2\mu\;6}^{\;a}\bigr)\;+
\\
&&\qquad e_a^\mu\;\bigl(
Q_{1\mu\;5}^{\;a}\,+\,Q_{2\mu\;6}^{\;a}\,-\,Q_{1\mu\;a}^{\;5}\,-\,Q_{2\mu\;a}^{\;6}\bigr)\bigr]\;+\nonumber\\
&&\lambda\,\bigl(L_{\;5}^5\,+\,L_{\;6}^6\bigr)\bigr\}\nonumber
\end{eqnarray}
where $\lambda =Tr\,((k\,k^*)^2)-1$. A straightforward but
lengthy computation shows that the Hilbert-Einstein action, for
unitary connections invariant under isometries and such that
$Tr_k\,T^A=0$ holds, is given by
\begin{eqnarray}
&& \cint r\;=\; \int \sqrt{g}\,d^4x\,\bigl[ -\frac 3
2\;R(e)\,+\,e^{-2\sigma}\,\bigl(2(\omega_{2\;a}^a)^2-
6\,(\omega_{2\;b}^a)^2- 4\,\omega_{2\;a}^a -
4\,\omega_{2\;a}^a\;\omega_{1\;6}^5\bigr)\nonumber\\
&&+\;6\;e^{-\sigma}\nabla_a\,(e^\sigma\,\nabla_a\sigma)\,+\,\lambda\,
\bigl( 2 (\partial_a\sigma)^2\,+\,e^{-2\sigma}\,
\bigl((\omega_{1\;6}^5)^2-2)\bigr)\bigr]
\end{eqnarray}
where $R(e)$ is the usual scalar curvature of $M_4$. The fields
$\omega_{2\;b}^a$ and $\omega_{1\;6}^5$ can be eliminated by their
equations of motion
\begin{eqnarray}
&& \omega_{2\;a}^a\;=\;\frac\lambda 2\;
\omega_{1\;6}^5\;=\;\frac{4}{1-\frac{8}{\lambda}} \nonumber\\
&& \omega_{2\;b}^a\;=\;0\;,\quad a\;\neq\;b .
\end{eqnarray}
Inserting (3.46) into (3.45) we get our final result
\begin{equation}
\cint r\;=\; \int
\sqrt{g}\;d^4x\,\bigl[ -\frac 3 2 \;R(e) +
2\,(3+\lambda)(\partial_z\sigma)^2 + c(\lambda)\;e^{-2\sigma}\bigr]
\end{equation}
where $c(\lambda) = \frac{\lambda(2\lambda-8)}{8-\lambda}$.

\noindent Using eq.~(3.22) and the definition of $\lambda$, one proves
that $\lambda\in [0,2]$. This implies that
$c(\lambda)\in\bigl[-\frac{4}{3}, 0\bigr]$ and it follows that the
potential is negative-definite.

\medskip

In reality, this is not the full story. We have only dealt, so far,
with the Dirac operator of the leptonic sector. In the standard model,
the quark sector must also be introduced. The Dirac operator of the
quark sector acts on the space of spinors
\begin{equation}
Q\;=\; \left( \begin{array}{c}
u_L \\ d_L \\ d_R \\ u_R \end{array} \right)
\end{equation}
and takes the form:
\begin{equation}
D\;=\;\left( \begin{array}{ccc}
\partial\!\!\!/ \;\otimes\,1_2\otimes\,1_3 &
k'\gamma_5\,e^{-\sigma}\;{0\choose 1}\enskip , & k''
\,e^{-\sigma}\,\gamma_5\;{-1\choose 0} \\
k^{'*} \,\gamma_5\,e^{-\sigma} (0,1)\enskip , & k^{''*}
\,e^{-\sigma}\,\gamma_5 (-1,0)\enskip , &
\partial\!\!\!/\,\otimes\,1_3
\end{array}\right)\;.
\end{equation}
Elements of the algebra $\a$ now have the form
\begin{equation}
a\;\to\;\left( \begin{array}{ccc}
(a_1)_{mn} \enskip &  & \\
                 & a_2\enskip & \\
                 &  & \bar a_2
\end{array} \right)\; .
\end{equation}
There is no increase in the number of independent components of
$\Omega_{\;B}^A$, because of the symmetry present in (3.50). The form
of the gravitational action in the quark sector will be the same as
(3.47), but now with different coefficients and with dependence on the
generation-mixing matrices $k'$ and $k''$ of the $d$ and $u$ quark
masses. The total action is of the form
\begin{equation}
\cint r\;=\;\cint (c_l\,r_l\;+\;c_q\,r_q)
\end{equation}
where $c_l$ and $c_q$ are arbitrary constants.

The total gravitational action, after eliminating the auxiliary
fields, is given by
\begin{equation}
\cint r\;=\;\int d^4x\,\bigl[ -\frac 1 2\;(3 c_l
+4c_q)R+\alpha(\nabla_a\sigma)^2 +\beta\;e^{-2\sigma}\bigr]
\end{equation}
where
\begin{eqnarray}
\alpha &=& \alpha (c_l,c_q,k_e,k_u,k_d)\nonumber\\
\beta &=& \beta (c_l,c_q,k_e,b_u,k_d)\nonumber
\end{eqnarray}
are coupling constants completely determined in terms of $c_l,c_q$ and
the electron and up and down quark generation-mixing matrices. We
choose the normalizations such that
\begin{eqnarray}
3 c_l\,+\,4 c_q &=& \frac{1}{k^2} \nonumber\\
\alpha\; &=& \frac 1 2
\end{eqnarray}
which can be solved for $c_l$ and $c_q$. Then $\beta$ is only a
function of $k_e$, $k_u$ and $k_d$. One would hope that $\beta$ will be
positive. However, since the gravitational action is
non-renormalizable, and in the absence of any understanding of quantum
non-commutative geometry, these coefficients do not have any physical
significance. The field $\sigma$, being the field whose $VEV$
determines the electroweak scale, plays the role of a link between the
gravitational sector and the low-energy sector and may provide a
signal of the non-commutative geometric structure of space time.

In a previous paper [8], two of the authors have studied the
low-energy effective potential and have shown that the field $\sigma$
acquires a well-determined $VEV$ at the quantum level, for limited
ranges of the top quark and Higgs masses. One of the solutions
obtained (corresponding to a top quark mass of $\sim$ 147 $Gev$) turns
out to correspond to a saddle point and is physically
unacceptable. The other solution obtained corresponds to a very heavy
Higgs mass and lies in the domain, where perturbation theory breaks
down, and the formula for the one-loop effective potential cannot be
trusted. One point which now is different from the starting point of
[8] is that $\beta$ was zero while now it is non-zero, in general.

This case was included in the analysis of Buchm\"uller and Busch [15],
who found an upper bound on the top quark mass of $\sim$ 100 $Gev$. This
bound is now experimentally excluded, signaling that nature lies
outside the perturbative domain. On the basis of the results found in
this paper and in [8, 15], we dare claim that if space-time has a
non-commutative structure responsible for the standard model then if
the top quark mass is in the suspected energy range around  $\sim$ 170
$Gev$ the
one-loop effective action cannot be trusted, and it is likely that the
Higgs mass is heavy. To get exact values, one would have to use the exact
effective potential which cannot be evaluated perturbatively.

\vfill\eject

\begin{appendix}

\section*{Appendix}

\renewcommand{\thesection}{A}    
\setcounter{equation}{0}         

\noindent \ub{{\bf Constraints equations}}

\medskip

\noindent Since the module $\tilde\Omega_D^1 \ca$ is not free, the connection
coefficients $\Omega_{\;B}^A$ are not arbitrary. In order to compute
the constraints, we take an arbitrary matrix of 1-forms
$\tilde\Omega_{\;B}^A$ and we use eq.~(2.39)
\begin{equation}
\Omega_{\;B}^A\;=\;\varepsilon^A (E_c)
\,\tilde\Omega_{\;D}^c\;\varepsilon_B (E^D)\,-\,d\varepsilon_B (E^A) .
\end{equation}
Comparing the matrix elements of these 1-forms we get the constraints,
\begin{eqnarray}
\omega_{1\mu\enskip 5,12}^{\quad a} &=&
\omega_{1\mu\enskip 5,22}^{\quad a}\enskip=\enskip
\omega_{1\mu\enskip 6,11}^{\quad a}\enskip=\enskip
\omega_{1\mu\enskip 6,21}^{\quad a}\enskip=\enskip0
\nonumber\\
\omega_{1\mu\enskip 5,11}^{\quad a} &=&
\omega_{1\mu\enskip 6,12}^{\quad a} \nonumber\\
\omega_{1\mu\enskip 5,21}^{\quad a} &=&
\omega_{1\mu\enskip 6,22}^{\quad a}\\
\tilde\omega_{2\enskip 5}^{\enskip a} &=&
\tilde\omega_{1\enskip 6}^{\enskip a}\enskip=\enskip 0 \nonumber\\
\tilde\omega_{1\enskip 5}^{\enskip a} &=&
\tilde\omega_{2\enskip 6}^{\enskip a} \nonumber
\end{eqnarray}
\begin{eqnarray}
\omega_{1\mu\enskip a,21}^{\quad 5} &=&
\omega_{1\mu\enskip a,22}^{\quad 5}\enskip=\enskip
\omega_{1\mu\enskip a,11}^{\quad 6}\enskip=\enskip
\omega_{1\mu\enskip a,12}^{\quad 6}\enskip =\enskip 0 \nonumber\\
\omega_{1\mu\enskip a,11}^{\quad 5} &=&
\omega_{1\mu\enskip a,21}^{\quad 6}\nonumber\\
\omega_{1\mu\enskip a,12}^{\quad 5} &=&
\omega_{1\mu\enskip a,22}^{\quad 6}\\
\omega_{2\enskip a}^{\enskip 5} &=&
\omega_{1\enskip a}^{\enskip 6}\enskip =\enskip 0\nonumber\\
\omega_{1\enskip a}^{\enskip 5} &=&
\omega_{2\enskip a}^{\enskip 6}\nonumber
\end{eqnarray}
\begin{eqnarray}
\omega_{1\mu\enskip 5,12}^{\quad 5} &=&
\omega_{1\mu\enskip 5,21}^{\quad 5}\enskip =\enskip
\omega_{1\mu\enskip 5,22}^{\quad 5}\enskip =\enskip 0\nonumber\\
\omega_{1\mu\enskip 6,11}^{\quad 5} &=&
\omega_{1\mu\enskip 6,21}^{\quad 5}\enskip =\enskip
\omega_{1\mu\enskip 6,22}^{\quad 5}\enskip =\enskip 0\nonumber\\
\omega_{1\mu\enskip 5,11}^{\quad 6} &=&
\omega_{1\mu\enskip 5,12}^{\quad 6}\enskip =\enskip
\omega_{1\mu\enskip 5,22}^{\quad 6}\enskip =\enskip 0\nonumber\\
\omega_{1\mu\enskip 6,11}^{\quad 6} &=&
\omega_{1\mu\enskip 6,12}^{\quad 6}\enskip =\enskip
\omega_{1\mu\enskip 6,21}^{\quad 6}\enskip =\enskip 0\nonumber\\
\omega_{1\mu\enskip 5,11}^{\quad 5} &=&
\omega_{1\mu\enskip 6,12}^{\quad 5}\enskip =\enskip
\omega_{1\mu\enskip 5,21}^{\quad 6}\enskip =\enskip
\omega_{1\mu\enskip 6,22}^{\quad 6}\phantom{Zeichn} \\
\omega_{2\enskip 5}^{\enskip 5} &=& - 1,\enskip
\tilde\omega_{2\enskip 5}^{\enskip 5}\enskip =\enskip 1 \nonumber\\
\omega_{2\enskip 6}^{\enskip 5} &=&
\omega_{1\enskip 5}^{\enskip 6}\enskip =\enskip
\omega_{1\enskip 6}^{\enskip 6}\enskip =\enskip 0\nonumber\\
\tilde\omega_{1\enskip 6}^{\enskip 5} &=&
\tilde\omega_{2\enskip 5}^{\enskip 6}\enskip =\enskip
\tilde\omega_{1\enskip 6}^{\enskip 6}\enskip =\enskip 0 \nonumber\\
\omega_{1\enskip 5}^{\enskip 5} &=&
\omega_{2\enskip 5}^{\enskip 6},\enskip
\tilde\omega_{1\enskip 5}^{\enskip  5}
\enskip =\enskip \tilde\omega_{2\enskip 6}^{\enskip 5}\nonumber\\
\omega_{1\enskip 6}^{\enskip 5}-1 &=&
\omega_{2\enskip 6}^{\enskip 6},\enskip
\tilde\omega_{2\enskip 6}^{\enskip 6} - 1
\enskip =\enskip  \tilde\omega_{1\enskip 5}^{\enskip 6}. \nonumber
\end{eqnarray}
At this point it is worth noting that there is another way of
computing the constraints. We consider all vanishing linear
combinations
\begin{equation}
\alpha_A\;E^A\;=\;0\;, \quad \alpha_A\,\in\,\a .
\end{equation}
This equation holds if and only if
\begin{eqnarray}
\alpha_a &=& 0\; , \quad a\;=\;1,\cdots, 4 \nonumber \\
\alpha_r &=&
\left( \begin{array}{ccc}
a_{r,11} & a_{r,12} & 0 \\
a_{r,21} & a_{r,22} & 0 \\
0        & 0        & 0
\end{array} \right)
\end{eqnarray}
together with the conditions
\begin{equation}
a_{5,11} \;+\;a_{6,12}\;=\;0\;,\quad a_{5,21}\;+\;a_{6,22}\;=\;0 .
\end{equation}
Then, we get constraints on the connection coefficients by imposing
\begin{equation}
\alpha_A\;E^A\;=\;0\;\Longrightarrow \;\nabla\,(\alpha_A\;E^A)\;=\;0 .
\end{equation}
Unfortunately this simpler way gives less constraints than eq.~(A.1),
because it gives only a minimal set of constraints and leaves
arbitrary coefficients which don't contribute to the connection.

\bigskip

\noindent\ub{{\bf Unitarity conditions}}

\medskip

\noindent Here, we give only the equations which are independent of
eqs.~(A.2) - (A.4)
\begin{eqnarray}
& & \omega_{1\mu\enskip b,ij}^{\quad a} \;=\;
- \omega_{1\mu\enskip a,ji}^{\quad b} \nonumber\\
& & \omega_{2\mu\enskip b}^{\quad a} \;=\;
 - \omega_{2\mu\enskip a}^{\quad b}\\
& & \omega_{1\enskip b}^{\enskip a} \;=\;
\tilde\omega_{1\enskip a}^{\enskip b}\;, \;
\omega_{2\enskip b}^{\enskip a} \;=\;
\tilde\omega_{2\enskip a}^{\enskip b}\phantom{Zeichnung} \nonumber
\end{eqnarray}
\begin{eqnarray}
& & 2\,\omega_{1\mu\enskip 5,11}^{\quad a} \;+\;
\omega_{1\mu\enskip a,11}^{\quad 5}\;=\;0 \nonumber \\
& & 2\,\omega_{1\mu\enskip 5,21}^{\quad a} \;+\;
\omega_{1\mu\enskip a,12}^{\quad 5} \;=\;0\nonumber \\
& & \omega_{2\mu\enskip 5}^{\enskip a} \;+\;
\omega_{2\mu\enskip a}^{\enskip 5} \;=\;
\omega_{2\mu\enskip 6}^{\enskip a}\;+\;
\omega_{2\mu\enskip a}^{\enskip 6}\;=\;0 \\
& & \omega_{1\enskip 5}^{\enskip a}\;=\;
\tilde\omega_{1\enskip a}^{\enskip 5}\;,\;
\omega_{2\enskip 5}^{\enskip a}\;=\;
\tilde\omega_{2\enskip a}^{\enskip 5}\nonumber \\
& & \omega_{1\enskip 6}^{\enskip a}\;=\;
\tilde\omega_{1\enskip a}^{\enskip 6}\;,\;
\omega_{2\enskip 6}^{\enskip a}\;=\;\tilde\omega_{2\enskip a}^{\enskip 6}
\nonumber \\
& & 2\,\tilde\omega_{1\enskip 5}^{\enskip a} \;=\; \omega_{1\enskip
  a}^{\enskip 5} \nonumber
\end{eqnarray}
\begin{eqnarray}
& & \omega_{1\mu\enskip 5}^{\enskip 5} \;=\;
\omega_{2\mu\enskip 5}^{\enskip 5} \;=\;
\omega_{2\mu\enskip 6}^{\enskip 6} \;=\; 0\nonumber \\
& & \omega_{2\mu\enskip 6}^{\enskip 5} \;+\;
\omega_{2\mu\enskip 5}^{\enskip 6}\;=\;0\\
& & 2\,\tilde\omega_{1\enskip 5}^{\enskip 5} \;=\;
\omega_{1\enskip 5}^{\enskip 5}\;, \quad
2\;+\; 2 \,\tilde\omega_{1\enskip 5}^{\enskip 6} \;=\;
\omega_{1\enskip 6}^{\enskip 5} \nonumber
\end{eqnarray}

\bigskip

\noindent\underbar{{\bf Conditions of vanishing torsion}}

\medskip

\noindent Taking eqs.~(A.2) - (A.4) and (A.9) - (A.11) into account the matrix
elements of the components of the torsion read,
\begin{eqnarray}
T_{11}^a &=& \gamma^{\mu\nu}\,(\partial_\mu\,e_\nu^a\,+\,e_\nu^b\,
\omega_{1\mu\enskip b}^{\enskip a}) \nonumber \\
T_{22}^a &=&
\gamma^{\mu\nu}\,(\partial_\mu\,e_\nu^a\,+\,e_\nu^b\,\omega_{2\mu\enskip
  b}^{\enskip a})\;+\; 2\,(k^*k-1)\;e^{-\sigma}\,
\tilde\omega_{1\enskip 5}^{\enskip a} \nonumber\\
T_{12}^a &=& k\,\gamma^\mu\,\gamma^5\;
\left( \begin{array}{c}
2\,\omega_{1\mu\enskip 5,11}^{\quad a}\,-\, e^{-\sigma}\, e_\mu^b\,
\omega_{1\enskip b}^{\enskip a} \\
2\,\omega_{1\mu\enskip 5,21}^{\enskip a}\,-\, e^{-\sigma}\, e_\mu^b\,
\omega_{2\enskip b}^{\enskip a}
\end{array} \right) \\
T_{21}^a &=& - k^*\,\gamma^\mu\,\gamma^4\, (\omega_{2\mu\enskip
  5}^{\enskip a}\,+\, e^{-\sigma}\, e_\mu^b\, \omega_{1\enskip
    a}^{\enskip b},\; \omega_{2\mu\enskip 6}^{\enskip a}\,+\,
  e^{-\sigma}\,e_\mu^b\,\omega_{2\enskip a}^{\enskip b}) \nonumber
\end{eqnarray}
\begin{eqnarray}
T_{11}^5 &=& e_5\,\cdot\,(-2\gamma^{\mu\nu}\,e_\nu^a)\,(
\omega_{1\mu\enskip 5,11}^{\quad a}, \omega_{1\mu\enskip 5,21}^{\quad
  a}) \nonumber \\
T_{22}^5 &=& -\gamma^{\mu\nu}\,e_\nu^a\,\omega_{2\mu\enskip
  5}^{\enskip a}\,+\, (k^*k-1)\;e^{-\sigma}\,\omega_{1\enskip
  5}^{\enskip 5}\nonumber \\
T_{12}^5 &=&
e_5\,\cdot\,k\,\gamma^\mu\,\gamma^5\,(\partial_\mu\,\sigma-2
e^{-\sigma} \,e_\mu^a\,\tilde\omega_{1\enskip 5}^{\enskip a}) \\
T_{21}^5 &=& -\,k^*\,\gamma^\mu\,\gamma^5\,(
\partial_\mu\,\sigma+e^{-\sigma}\, e_\mu^a\,\omega_{1\enskip 5}^a,\;
e^{-\sigma}\,e_\mu^a\, \omega_{2\enskip 5}^{\enskip
  a}\,+\,\omega_{2\mu\enskip 6}^{\enskip 5})\nonumber
\end{eqnarray}
\begin{eqnarray}
T_{11}^6 &=& e_6\,\cdot\, (-2\gamma^{\mu\nu}\,e_\nu^a)\,
(\omega_{1\mu\enskip 5,11}^{\quad a}, \; \omega_{1\mu\enskip
  5,21}^{\quad a}) \nonumber \\
T_{22}^6 &=& -\gamma^{\mu\nu}\,e_\nu^a\,\omega_{2\mu\enskip
  6}^{\enskip a} \,+\, (k^*k-1)\,e^{-\sigma}\,(\omega_{1\enskip
  6}^{\enskip 5} - 1)\nonumber \\
T_{12}^6 &=& e_6\,\cdot\,k\,\gamma^\mu\,\gamma^5\,(\partial_\mu \sigma
- 2 e^{-\sigma}\,e_\mu^a\, \tilde\omega_{1\enskip 5}^{\enskip a}) \\
T_{21}^6 &=& -k^*\,\gamma^\mu\,\gamma^5\,(e^{-\sigma}\, e_\mu^a\,
\omega_{1\enskip 6}^{\enskip a} - \omega_{2\mu\enskip 6}^{\enskip
  5},\; \partial_\mu \sigma + e^{-\sigma}\, e_\mu^a\, \omega_{2\enskip
  6}^{\enskip a})\; . \nonumber
\end{eqnarray}
Then, imposing $Tr_k T^A=0$ and using eq.~(3.22) we get the following
equations
\begin{eqnarray}
& & \gamma^{\mu\nu}\,(\partial_\mu\,e_\nu^a +
e_\nu^b\,\omega_{1\mu\enskip b}^{\quad
a})\;=\;\gamma^{\mu\nu}\,(\partial_\mu\,e_\nu^a +
e_\nu^b\,\omega_{2\mu\enskip b}^{\quad a})\;=\;0\nonumber \\
& & 2\,\omega_{1\mu\enskip 5,11}^{\quad a} -
e^{-\sigma}\,e_\mu^b\,\omega_{1\enskip b}^{\enskip a}\;=\;
2\,\omega_{1\mu\enskip 5,21}^{\quad a} -
e^{-\sigma}\,e_\mu^b\,\omega_{2\enskip b}^{\enskip a}\;=\;0 \\
& & \omega_{2\mu\enskip 5}^{\enskip a} +
e^{-\sigma}\,e_\mu^b\,\omega_{1\enskip a}^{\enskip b}\;=\;
\omega_{2\mu\enskip 6}^{\quad a} + e^{-\sigma}\,e_\mu^b\,
\omega_{2\enskip a}^{\enskip b}\;=\;0\nonumber
\end{eqnarray}
\begin{eqnarray}
& & \gamma^{\mu\nu}\,e_\nu^a\,\omega_{1\mu\enskip 5,11}^{\quad a}\;=\;
\gamma^{\mu\nu}\,e_\nu^a\,\omega_{1\mu\enskip 5,21}^{\quad a}\;=\;
\gamma^{\mu\nu}\,e_\nu^a\,\omega_{2\mu\enskip 5}^{\enskip a}
\;=\;0\nonumber \\
& & \gamma^{\mu\nu}\,e_\nu^a\,\omega_{2\mu\enskip 6}^{\quad a}
\;=\;0\nonumber\\
& & \partial_\mu\,\sigma - 2\,e^{-\sigma}\,e_\mu^a\,
\tilde\omega_{1\enskip 5}^{\enskip a}\;=\;0 \\
& & \partial_{\mu}\,\sigma + e^{-\sigma}\,e_\mu^a\,\omega_{1\enskip
  5}^{\enskip a}\;=\;0\nonumber \\
& & \partial_\mu\,\sigma + e^{-\sigma}\,e_\mu^a\, \omega_{2\enskip
  6}^{\enskip a}\;=\;0\nonumber \\
& & e^{-\sigma}\,e_\mu^a\,\omega_{2\enskip 5}^{\enskip a} +
\omega_{2\mu\enskip 6}^{\quad 5}\;=\; e_\mu^a\,e^{-\sigma}\,
\omega_{1\enskip 6}^{\enskip a} - \omega_{2\mu\enskip 6}^{\quad
  5}\;=\;0.\nonumber
\end{eqnarray}
If we impose $T^A=0$ we get the additional equations
\begin{equation}
\tilde\omega_{1\enskip 5}^{\enskip a}\;=\;\omega_{1\enskip 5}^{\enskip
  5}\;=\;\omega_{1\enskip 6}^{\enskip 5} - 1\;=\; 0,
\end{equation}
and this implies together with eq.~(A.16)
\begin{equation}
\partial_\mu\,\sigma\;=\;0 .
\end{equation}
Thus, if we want the $\sigma$-field to be dynamical we should only
require \ $Tr_k\,T^A = 0$.

\bigskip

\noindent\ub{{\bf Invariance under isometries}}

\medskip

\noindent The new constraints coming from the invariance of the
connection under isometries are,
\begin{eqnarray}
& & \omega_{1\mu\enskip b,12}^{\quad a} \;=\; 0 \qquad
\omega_{1\enskip b}^{\enskip a}\;=\; 0 \\
&& \omega_{1\mu\enskip 5,11}^{\quad a}\;=\; \omega_{2\mu\enskip
  5}^{\quad a}\;=\;0, \enskip \omega_{2\enskip 5}^{\enskip a}\;=\;
\omega_{1\enskip 6}^{\enskip a}\;=\; 0 \\
&& \omega_{2\mu\enskip 6}^{\quad 5}\;=\;0, \quad \omega_{1\enskip
  5}^{\enskip 5}\;=\; \omega_{1\enskip 6}^{\enskip a}\;=\;0.
\end{eqnarray}

\medskip

\noindent\ub{{\bf The components of the curvature}}

\medskip

\noindent We give explicit formulas for the curvature components.
\begin{eqnarray}
&& R_{\mu\nu\enskip B}^{(1)\enskip A}\;=\;\frac 1 2 \; \bigl(
\partial_\mu\,\omega_{1\nu\enskip B}^{\quad A} + \omega_{1\mu\enskip
  C}^{\quad A}\, \omega_{1\nu\enskip B}^{\quad C} -
(\mu\leftrightarrow \nu)\bigr) \nonumber \\
&& R_{\mu\nu\enskip B}^{(2)\enskip A}\;=\;\frac 1 2 \;\bigl(
\partial_\mu\,\omega_{2\nu\enskip B}^{\quad A} + \omega_{2\mu\enskip
  C}^{\quad A}\,\omega_{2\nu\enskip B}^{\quad C} -
(\mu\leftrightarrow\nu) \bigr)\\
&& \nonumber \\
&& P_{\mu\enskip B}^{\enskip A}\;=\;
\left( \begin{array}{c}
\partial_\mu\,\omega_{1\enskip B}^{\enskip A} + \omega_{1\mu\enskip
  B,12}^{\quad A} \\
\partial_\mu\,\omega_{2\enskip B}^{\enskip A} + \omega_{1\mu\enskip
  B,22}^{\quad A} - \omega_{2\mu\enskip B}^{\quad A}
\end{array} \right)\;+\;
\omega_{1\mu\enskip C}^{\quad A}\,\omega_B^C -
\omega_C^A\,\omega_{2\mu\enskip B}^{\quad C}\nonumber \\
&& Q_{\mu\enskip B}^{\enskip A}\;=\;
\left( \begin{array}{c}
\partial_\mu\,\tilde\omega_{1\enskip B}^{\enskip A} -
\omega_{1\mu\enskip B,21}^{\quad A} \\
\partial_\mu\,\tilde\omega_{2\enskip B}^{\enskip A} -
\omega_{1\mu\enskip B,22}^{\quad A} + \omega_{2\mu\enskip B}^{\quad A}
\end{array} \right)^T \;-\;
\tilde\omega_C^A\,\omega_{1\mu\enskip B}^{\quad C} +
\omega_{2\mu\enskip C}^{\quad A}\, \tilde\omega_B^C
\end{eqnarray}
\begin{equation}
L_B^A\;=\;e^{-2\sigma}\,(\omega_{2\enskip B}^{\enskip A} +
\tilde\omega_{2\enskip B}^{\enskip A} + \tilde\omega_C^A\,\omega_B^C).
\end{equation}

\bigskip

\noindent\ub{{\bf The components of the Ricci tensor}}

\medskip

\noindent The Ricci tensor is given by \ Ric$_B = (P_D^{(1)} -
P_D^{(0)}) (\tilde\varepsilon_A\,R_B^A)$, see eq.~(2.35). One finds
\begin{eqnarray}
&& {\rm Ric}_{B,11}\;=\;(e_a^\mu\,\gamma^\nu - e_a^\nu\,\gamma^\mu)\,
R_{\mu\nu\;B}^{(1)\,a} + e^{-\sigma}\,\gamma^\mu\,e_r\,Q_{\mu\;B}^r
\nonumber \\
&& {\rm Ric}_{B,22}\;=\;(e_a^\mu\,\gamma^\nu - e_a^\nu\,\gamma^\mu)\,
R_{\mu\nu\;B}^{(2)\,a} - e^{-\sigma}\,\gamma^\mu\;\frac 1 2\; e_r^\top\,
P_{\mu\;B}^r \\
&& {\rm Ric}_{B,12}\;=\;k\,\gamma^5\,\bigl(
e_a^\mu\,e^{-\sigma}\,P_{\mu\;B}^a -
Tr\,\bigl((k^*k)^2-1\bigr)\;e_r\,L_{\;B}^r\bigr) \nonumber \\
&& {\rm
  Ric}_{B,21}\;=\;k^*\,\gamma^5\,e^{-\sigma}\,e_a^\mu\,Q_{\mu\;B}^a
. \nonumber
\end{eqnarray}

\end{appendix}

\vfill\eject

\section*{References}

\medskip

\begin{enumerate}

\item A. Connes and J. Lott, Nucl.~Physics Proc.~Supp.~{\it 18B}
  (1990) 295.\hfill\break
  A. Connes and J. Lott, Proceedings of 1991 Carg\`ese Summer
  Conference, ed. by J.~Fr\"ohlich et al. (Plenum, New York 1992).
\item A. Connes, Publ.~Math.~IHES {\it 62} (1983) 44; \hfill\break
 A. Connes, ``Non-Commutative Geometry'' Academic Press, (1994).
\item A. Chamseddine, G. Felder and J. Fr\"ohlich,
  Comm.~Math.~Phys.~{\it 155} (1993) 109.
\item A. Connes, ``Non-Commutative Geometry and Physics'', Les Houches
  Lecture Notes (1992).
\item D. Kastler, ``The Dirac operator and gravitation'',
  Comm.~Math.~Phys.~{\it 166} (1995) 633.
\item W. Kalau and M. Waltze, Mainz University Preprint MZ-TH/93-38.
\item A. Chamseddine and J. Fr\"ohlich, in preparation.
\item A. Chamseddine and J. Fr\"ohlich, Phys.~Lett.~{\it 314 B} (1993)
  308.
\item A. Chamseddine and J. Fr\"ohlich, ``Some elements of
  non-commutative and space-time geometry'', Yang Festschrift,
  ed. S.T.Yau.
\item R. Swan, Trans.~Amer.~Math.~Soc.~{\it 105} (1962) 264.
\item B. Bleile, diploma thesis, ETH 1993.
\item First reference in 1.
\item P. Hilton and Y.-C. Wu., ``A course in modern algebra'', John
  Wiley, New York 1974.
\item A. Chamseddine, J. Fr\"ohlich and O. Grandjean, in preparation.
\item W. Buchm\"uller and C. Busch, Nucl.~Phys.~{\it B 349} (1991) 71.
\end{enumerate}

\end{document}